\documentclass[notitlepage,twocolumn,prl,showpacs,amsmath,amssymb]{revtex4-1}
\usepackage{amssymb}
\usepackage{graphicx}
\usepackage{bm}
\usepackage{epstopdf}
\usepackage[super]{nth}
\usepackage{epsfig}
\usepackage{dsfont}
\usepackage{bigints}
\usepackage{amsmath}
\usepackage{relsize}
\usepackage{relsize}
\usepackage{amsmath}
\usepackage{accents}
\newlength{\dhatheight}

\newcommand{\unit}{1\!\!1}
\newcommand{\s}[1]{\scriptscriptstyle{#1}}

\begin{document}

\title{Signatures of causality and determinism in a quantum theory of events}

\author{Aditya Iyer$^1$}
\email{aditya.iyer@physics.ox.ac.uk}
\author{Eduardo O. Dias$^2$}%
\email{eduardo.dias@ufpe.br}
\author{Vlatko Vedral$^{1,3,4}$}
\email{vlatko.vedral@physics.ox.ac.uk}
\affiliation{%
$^1$Clarendon Laboratory, Department of Physics, University of Oxford, Oxford OX1 3PU, United Kingdom
}%
\affiliation{%
$^2$Departamento de Fisica, Universidade Federal de Pernambuco, Recife, Pernambuco 50670-901, Brazil
}%
\affiliation{
$^3$Centre for Quantum Technologies, National University of Singapore, Block S15, 3 Science Drive 2, Singapore
}%
\affiliation{$^4$Department of Physics, National University of Singapore, 2 Science Drive 3, Singapore
}
\begin{abstract}
By representing an event as the joint state of a detector-timer couple that interact with a system, we recover the familiar tensor product structure, used to describe spatially separated systems, in the context of timelike events. Furthermore, with this approach, we extend the superposition principle to the moment of occurrence of events. We then outline quantum signatures of causality that manifest through coherence in the detector state and correlation functions of time operators. Finally, we expand the scope of quantum information theoretic measures of state discrimination and information content, commonly used to characterize spatially separated systems, to events in spacetime. For causally connected events, we illustrate a deterministic relationship between events (akin to spatially entangled physical systems) where observing a previous event (one subsystem), enables us to delineate a later event (the other subsystem).
\end{abstract}


\maketitle

\textit{Introduction}--- In standard quantum mechanics (QM), position is an operator while time is an independent parameter. This asymmetry between space and time is further manifested in the absence of a tensor product description of temporally separated states. Recall that this tensor product structure for spatially separated systems, along with superposition principle, yields the property of entanglement. Despite lacking an analogous mathematical construction in the treatment of time, QM predicts non-classical temporal correlations~\cite{Leg,Paz,See,Vedral3,Now}. These correlations ---  dubbed entanglement in time --- concern timelike events, referring to measurements performed at different times on the same physical system.

Since temporal quantum correlations can also be accessed via Bell-like inequalities~\cite{Vedral3}, one can expect QM to incorporate a tensor product structure for both spacelike events (SLE) and causally connected timelike events (TLE). Besides Ref.~\cite{Dias}, proposals for a more symmetrical approach to QM~\cite{Re1,Re2,Re3,Re4,Re5,Re6,Re7,Re8, Fink,Isham,Now,Fit,Cot,Dias} assume ``instantaneous'' measurements, where unlike position, events do not present any time-measurement uncertainty \cite{footnote1}. In addition, none of these formalisms exploit the tensor product structure to study quantum information theoretic aspects of events, especially the causally connected ones.

Here we utilize elements of Ref.~\cite{Dias} to recover a tensor product for TLE. Ref.~\cite{Dias} defines states of events by using the Hilbert spaces of the measured system, detectors and timers. Here however, we obtain an equivalent description for SLE and TLE by tracing out the system under measurement and focusing only on the detectors and timers.  As a result, in departure from previous approaches (including~\cite{Dias}), the theoretical framework used here allow us to extend quantum information (QI) tools \cite{book1,book2} to events as follows.

Firstly, we will revisit measures of coherence, applied in the context of events, to serve as an indicator of causality when one does not have information about the moment when the events occurred. We will also propose a correlation function of time operators that serves as an indicator of causality. While Ref.~\cite{Fit} outlines a measure of causality, their proposed extension of QM to include a tensor products description of temporally separated events --- the so-called pseudo-density matrices (PDM) --- results in states with negative eigenvalues. Here, in contrast, we remain within the scope of traditional QM, with the simple addition of timers and detectors to our description. Secondly, we extend measures of classical correlation~\cite{Vedral2} and state discrimination criteria~\cite{Helstrom,Holevo} to events to quantify, for causally connected events, how much information we know about the outcomes of future events. In specific cases, we will illustrate a deterministic feature of two TLE.

\textit{Describing events in QM}--- Consider an event to be a measurement carried out by a detector $\cal D$ (micro/macroscopic), coupled with a timer $\cal T$, on a system $\cal S$. Let ${\hat H}_{\cal S}$ be the Hamiltonian of $\cal S$ associated with the degrees of freedom under measurement. In SLE, $\cal S$ is composed of two systems $S_A$ and $S_B$ (entangled or otherwise) that are measured once, independently. We denote the measurement on ${\cal S}_A$ (${\cal S}_B$) by event $A$ (event $B$). On the other hand, in TLE, we investigate causally connected events such that a single system $\cal S$ is measured twice. We call the first measurement ``event $A$'' and the second ``event $B$.'' In both SLE and TLE, the observables of $\cal S$ are ${\hat \alpha_j}=\sum_{\alpha_j}\alpha_j|\alpha_j\rangle_{{\cal S}_j}{\langle {\alpha}_j|}$, with $j=A,B$. 

To record the moment of occurrence of an event, one can consider a Salecker-Wigner-Peres-like timer $\cal T$~\cite{Pe}. Like Ref.~\cite{Dias}, instead of being coupled with the system $\cal S$, as in Ref.~\cite{Pe}, here $\cal T$ interacts with the detector in such a way $\cal T$ stops running when $\cal D$ measures $ \cal S$. Thus, while $\cal D$ ``is'' in a ready state $ |0 \rangle_{\cal D}$, $ \cal T $ evolves as an ideal clock by the Hamiltonian ${\cal H}_{\cal T}$. Consider that the time scale of the interaction between the timer and detector is significantly smaller than the characteristic time $\Delta t_{{\cal S}{\cal D}}$ of the measurement of $\cal S$ by $\cal D$. An interaction between $\cal T$ and $\cal D$ that models this physical picture is ${\hat V}_{\cal TD}=\unit_{\cal S} \otimes {\hat H}_{\cal T}\otimes |0\rangle_{ \cal D}{\langle} 0|$~\cite{Pe}, and the total Hamiltonian is ${\hat H}={\hat H}_{\cal S}\otimes\unit_{\cal TD}+{\hat V}_{\cal SD}+  {\hat V}_{\cal TD}$, where ${\hat V}_{\cal SD}$ is the interaction potential between $\cal S$ and $\cal D$. Also, let the timer's observable $\hat T$ be such that ${\hat T}|t\rangle_{\cal T}=t|t\rangle_{\cal T}$ and $[{\hat T},{\hat H}_{\cal T}]=i\hbar$. 

By considering a pair of $\cal TD$ to perform two measurements, regardless of the type of events, the initial condition of the full state is
$|\psi(t_0)\rangle=|\varphi(t_0)\rangle_{\cal S}\otimes |t_0,0\rangle_A\otimes |t_0,0\rangle_B.$
Here, $|\varphi(t_0)\rangle_{\cal S}$ is the initial state of $\cal S$ and $|t_0,0\rangle_j=|t_0\rangle_{{\cal T}_j}\otimes |0\rangle_{{\cal D}_j}$ ($j=A,B$), with $|t_0\rangle_{{\cal T}_j}$ being the initial state of the timer ${\cal T}_j$ and $|0\rangle_{{\cal D}_j}$ the ready state of the detector ${\cal D}_j$. To compute $|\psi(t)\rangle=\exp\{-i(t-t_0){\hat H}/\hbar\}|\psi(t_0)\rangle$, let us break up the Schr\"odinger evolution into infinitesimal steps $\delta t\ll \Delta t_{ \cal SD}$, so that the eigenvalues of $\cal T$ (denoted $t$) in this discrete evolution are $t_0, t_{\s (1)}, \hdots, t_{\s (N)}, \hdots$, with $\delta t=t_{\s (k+1)}-t_{\s (k)}$.

To describe two SLE, we model two independent measurements of two systems ${\cal S}_A$ and ${\cal S}_B$ such that in the first step $t_{\s (1)}$, the full initial state splits into four branches given by
\begin{eqnarray}
\label{inf1}
&&|\psi(t_{\s (1)})\rangle
=\sqrt{q^A_{\s (1)}q^B_{\s (1)}}~{\rm e}^{i(\varphi^A_{(1)}+\varphi^B_{(1)})}|\varphi(t_{\s (1)})\rangle_{\cal S}|t_{\s (1)},0\rangle_{A}|t_{\s (1)},0\rangle_{B}\nonumber\\
&& +\sqrt{\delta p^A_{\s (1)}~q^B_{\s (1)}}~{\rm e}^{i\varphi^B_{(1)}}\sum_{\alpha_A}{\hat M}_{\alpha_A}^A|\varphi(t_{\s (1)})\rangle_{\cal S}~|t_{\s (1)},\alpha_A\rangle_{A}|t_{\s (1)},0\rangle_{B}\nonumber\\
&&+\sqrt{q^A_{\s (1)}~\delta p^B_{\s (1)}}~{\rm e}^{i\varphi^A_{(1)}}\sum_{\alpha_B}{\hat M}^B_{\alpha_B}|\varphi(t_{\s (1)})\rangle_{\cal S}~|t_{\s (1)},0\rangle_{A}|t_{\s (1)},\alpha_B\rangle_{B}\nonumber\\
&&+\sqrt{\delta p^A_{\s (1)}\delta p^B
_{\s (1)}}\sum_{\alpha_A,\alpha_B} {\hat M}_{\alpha_A,\alpha_B}^{AB}|\varphi(t_{\s (1)})\rangle_{\cal S}|t_{\s (1)},\alpha_A\rangle_{A}|t_{\s (1)},\alpha_B\rangle_{B},\nonumber\\
\end{eqnarray}
where $\delta p^j
_{\s (1)}\ll 1$ ($q^j_{\s (1)}=1-\delta p^j_{\s (1)}$) is the probability of ${\cal S}_j$ being (not being) measured in the interval $(t_0,t_{\s (1)}]$. $\varphi^j_{\s (1)}$ is the phase associated with the undetected branch of ${\cal S}_j$. Here ${\hat M}_{\alpha_A}^A={\hat M}_{\alpha_A}\otimes \unit_B$, ${\hat M}_{\alpha_B}^B=\unit_A\otimes {\hat M}_{\alpha_B}$, ${\hat M}_{\alpha_A,\alpha_B}^{AB}={\hat M}_{\alpha_A}\otimes{\hat M}_{\alpha_B}$, with ${\hat M}_{\alpha_j}=|\alpha_j\rangle_{j}{\langle}\alpha_j|$. Also, $|\varphi(t_{\s(1)})\rangle_{\cal S}={\hat U}_{{\cal S}}(t_{\s (1)},t_0)|\varphi(t_0)\rangle_{\cal S}$ where
${\hat U}_{\cal S}(t_{\s (1)},t_0)=\exp \{-i(t_{\s(1)}-t_0){\hat H}_{\cal S}/\hbar\}$. For simplicity we consider that the detectors interact equally with both systems and also with each state $|\alpha_j \rangle_{{\cal S}_j}$. 

Inspecting Eq.~(\ref{inf1}), we note that in the first branch the detectors do not record any information about ${\cal S}={\cal S}_A+{\cal S}_B$ with a high probability $q^A_{\s (1)}q^B_{\s (1)}=(1-\delta p^A_{\s (1)}) (1- \delta p^B_{\s (1)})$. Thus, the timers will continue to evolve as an ideal quantum clock. In the second branch, with probability $\delta p^A_{\s (1)}(1-\delta p^B
_{\s (1)})$, ${\cal D}_A$ measures ${\cal S}_A$ and  ${\cal D}_B$ does not measure ${\cal S}_B$, thus ${\cal T}_A$ stops running and records the instant of the event $A$, $t_{\s(1)}$. In the third branch, the opposite happens. In the last branch both systems are measured in the interval $(t_{\s(0)},t_{\s (1)}]$ and both timers register $t_{\s(1)}$. Note that two independent Stern-Gerlach's (SG's) devices, one for each particle ${\cal S}_A$ and ${\cal S}_B$, fits the description of Eq.~(\ref{inf1}) \cite{footnote2}. 

We are interested in $|\psi(t)\rangle$ from the moment we can guarantee both ${\cal S}_A$ and ${\cal S}_B$ have been measured. Thus, we evaluate $|\psi(t_{(N)})\rangle$ when $t_{\s (N)}\gtrapprox t_0 + t_{{\rm a}.{\rm t}.}+\Delta t_{\cal SD}$, where $t_{{\rm a}.{\rm t}.}$ is the typical arrival time at the detector. Here, in the continuous limit, WLOG we take $t_{\s (N)}\rightarrow \infty$ [$|\psi_e\rangle\equiv|\psi(t_{\s (N)}\rightarrow \infty)\rangle$] to ensure that the system has completed its interaction with the detector. Note that, we do not invoke environment induced decoherence and consider the complete unitary description of the system-detectors-timers' interactions. Now,
\begin{eqnarray}\label{psie}
|\psi_\varepsilon\rangle=\sum_{\alpha_A,\alpha_B}\int_{t_0}^\infty\int_{t_0}^\infty~dt_A dt_B~|\Phi_\varepsilon(t_A,\alpha_A;t_B,\alpha_B)\rangle_{\cal S}\nonumber\\
\otimes|t_A,\alpha_A\rangle_{A}~|t_B,\alpha_B\rangle_{B},~~
\end{eqnarray}
where $\varepsilon$=SL labels spacelike events. As our interest lies in measurement outcomes recorded by ${\cal T}_A{\cal D}_A$ and ${\cal T}_B{\cal D}_B$, we trace out ${\cal H}_{\cal S}$, yielding
\begin{eqnarray}\label{rhoe}
{\hat \rho}_\varepsilon=\sum_{\alpha_A,\alpha_B}~\int_{t_0}^\infty\int_{t_0}^\infty~dt_A dt_B\sum_{{\alpha}_A^{\prime},{\alpha}_B^{\prime}}~\int_{t_0}^\infty\int_{t_0}^\infty~dt'_A dt'_B\nonumber\\
{\rm Tr}_{\cal S}\Big\{|\Phi_\varepsilon(t_A,\alpha_A;t_B,\alpha_B)\rangle_{\cal S}\langle\Phi_\varepsilon(t'_A,\alpha'_A;t'_B,\alpha'_B)|\Big\}\nonumber\\
|t_A,\alpha_A\rangle_{A}~|t_B,\alpha_B\rangle_{B}~{_A{\langle}}t'_A,\alpha'_A|~{_B{\langle}}t'_B,\alpha'_B|,~~~~
\end{eqnarray}
which describes the events $A$ and $B$ with
\begin{eqnarray}
\label{SL}
&&|\Phi_{{\rm S}{\rm L}}(t_A,\alpha_A;t_B,\alpha_B)\rangle_{\cal S}
=\chi_{{\rm SL}}(t_A,t_B)~\nonumber\\
&&\Big[\Theta(t_B,t_A){\hat U}_{{\cal S}}(t_{\s(N)},t_B){\hat M}_{\alpha_B}^B~{\hat U}_{{\cal S}}(t_B,t_A){\hat M}_{\alpha_A}^A{\hat U}_{\cal S}(t_{A},t_0)\nonumber\\
&&+\Theta(t_A,t_B){\hat U}_{{\cal S}}(t_{\s (N)},t_A)~{\hat M}_{\alpha_A}^A{\hat U}_{{\cal S}}(t_A,t_B){\hat M}_{\alpha_B}^B{\hat U}_{\cal S}(t_{B},t_0)\nonumber\\
&&+\delta(t_B-t_A)~{\hat U}_{{\cal S}}(t_{\s (N)},t_A)~{\hat M}_{\alpha_A,\alpha_B}^{AB}~{\hat U}_{\cal S}(t_{A},t_0)\Big]|\varphi(t_0)\rangle_{\cal S}.\nonumber\\
\end{eqnarray}
Here $t_{\s(N)}\rightarrow \infty$, $\chi_{\rm {SL}}(t_A,t_B)=\chi_{A}(t_A)\chi_B(t_B)$, $\chi_j(t_j)$ is the probability density amplitude for ${\cal S}_j$ to be measured at the instant $t_j$, regardless of the outcome $\alpha_j$, and $\Theta(t_x,t_y)=1$ for $t_x>t_y$, or $0$ otherwise. Note, in general situations, $\chi_{\rm {SL}}(t_A,t_B)$ is non-separable. Inspecting Eqs.~(\ref{rhoe}) and (\ref{SL}), we observe a superposition of measurements of ${\cal S}_A$ and ${\cal S}_B$ in an indefinite order. In the second (third) line of Eq.~(\ref{SL}), ${\cal S}_A$ (${\cal S}_B$) is measured first. In the last branch, ${\cal S}_A$ and ${\cal S}_B$ is measured ``simultaneously'' in accordance with the coarse-grained nature of the evolution.

In Eq.~(\ref{SL}), $\chi_j(t_j)$ is the continuous limit of the amplitude
$
\tilde{\chi}_j(t_{\s (k)},t_{\s (k-1)})= ({\delta p^j_{\s (k)}})^{1/2}~\prod_{\ell=1}^{k-1}(1-\delta p^j_{\s (\ell)})^{1/2}~{\exp}({i\varphi^j_{\s (\ell)}}),
$
whose modulus squared is the probability of ${\cal S}_j$ being measured in $(t_{\s(k-1)},t_{\s (k)}]$. Notice the outcome of the product operator represents the probability amplitude for ${\cal S}_j$ to not be measured in  $(t_0,t_{\s(k-1)}]$. To obtain the continuous limit $\chi_j(t_j)$, we can use the relation $|\chi_j(t_{\s(k)})|^2~dt=|\tilde{\chi}_j(t_{(k+1)},t_{\s(k)})|^2$ (Appendix A shows an example). Eqs.~(\ref{rhoe}) and (\ref{SL}) could also describe polarization measurements (and the instant of their detection) of spontaneous decays. 

Finally, one can verify that ${\hat \rho}_\varepsilon$ of Eq.~(\ref{rhoe}) can also be applied to TLE ($\varepsilon=$TL), specifically, in the situation where a single system $\cal S$ is measured twice~\cite{Dias}. Assuming the second measurement (event $B$) begins soon after the first measurement (event $A$), we have
\begin{eqnarray}
\label{TL}
&&|\Phi_{{\rm T}{\rm L}}(t_A,\alpha_A;t_B,\alpha_B)\rangle_{\cal S}=\chi_{\rm {TL}}(t_A,t_B)\Theta(t_B>t_A)\nonumber\\
&&{\hat U}_{{\cal S}}(t_{\s (N)},t_B){\hat M}_{\alpha_B}{\hat U}_{{\cal S}}(t_B,t_A){\hat M}_{\alpha_A}{\hat U}_{\cal S}(t_{A},t_0)|\varphi(t_0)\rangle_{\cal S},~~~~\end{eqnarray}
where $\chi_{\rm {TL}}(t_A,t_B)=\chi_B(t_B|t_A)~\chi_A(t_A)$, with $\chi_A(t_A)$ as previously discussed. $\chi_B(t_B|t_A)$ is the probability amplitude for the second event to take place at time $t_B$, given the first measurement happens at $t_A$. Analogous to SLE, Eqs.~(\ref{rhoe}) and (\ref{TL}) also describe the experiment involving Stern-Gerlach devices, as well as the spontaneous decay discussed above. However, for TLE, we should have two consecutive SG setups (or beam splitters) measuring the same system $\cal S$. In the simpler case where the uncertainty in detection time is  negligible ($\chi_j$ is a Dirac delta function), the timers can be neglected and we recover the formalism of super-density operators~\cite{Cot}.

\textit{Signatures of causality in QM}--- First, for succinctness let us consider the aforementioned simpler case, and hence remove the timers. Thus, we proceed to juxtapose SLE and TLE with only the information gleaned from ${\cal D}_A$ and ${\cal D}_B$.  For two SLE, Eq.~(\ref{rhoe}) becomes
\begin{eqnarray}
\label{rhoSL}
{\hat \rho}_{{\rm S}{\rm L}}=\sum_{\alpha_A.\alpha_B}{\langle}\alpha_A,\alpha_B|{\hat \rho}_{\cal S}|\alpha_A,\alpha_B\rangle
|\alpha_A,\alpha_B\rangle_{AB}{\langle}\alpha_A,\alpha_B|,~~~~~
\end{eqnarray}
where $~|\alpha_A,\alpha_B\rangle=|\alpha_A\rangle_{{\cal S}_A}|\alpha_B\rangle_{{\cal S}_B}$. Henceforth, we drop the label ${\cal S}_j$ in $|\alpha_j\rangle_{{\cal S}_j}$. In the absence of timers, the detector could be a simple two-level atom prepared in the ground state, positioned in an arm of the SG apparatus. 

Similarly, utilizing the same picture for two TLE, and allowing the system to evolve between measurements via the unitary operation ${\hat U}={\hat U}_{\cal S}$, Eq.~(\ref{rhoe}) becomes
\begin{eqnarray}
\label{rhoTL}
{\hat \rho}_{{\rm T}{\rm L}}=\sum_{\alpha_A,\alpha'_A}\sum_{\alpha_B}\langle \alpha_B|{\hat U}|\alpha_A\rangle\langle \alpha_A|{\hat \rho}_{\cal S}|\alpha'_A\rangle\langle \alpha'_A|{\hat U}^\dagger |\alpha_B\rangle\nonumber\\
|\alpha_A,\alpha_B\rangle_{AB}{\langle}\alpha'_A,\alpha_B|.~~
\end{eqnarray}
From Eqs.~\eqref{rhoSL} and~\eqref{rhoTL}, the distinction between SLE and TLE becomes apparent: ${\hat \rho}_{{\rm S}{\rm L}}$ is diagonal in the eigenbasis of the measured observable ${\hat A}_j$, whereas ${\hat \rho}_{{\rm T}{\rm L}}$ contains off-diagonal terms associated with the Hilbert subspace of ${\cal D}_{A}$. Obviously, for ${\hat \rho}_{{\rm T}{\rm L}}$ to maintain its coherence, one should prevent the detectors from decoherence. We are led to the striking conclusion that by simply analyzing the state of the detector (without any information of when and how the events happened), the presence of coherence in ${\hat \rho}_\varepsilon$ indicates a causal relation between the events. Note that it is possible to choose a unitary ${\hat U}$ that destroys these coherence terms in the case of TLE, thereby rendering SLE and TLE indistinguishable. In Ref.~\cite{Fit}, the trace norm of the PDM reveals a causal relation between the events. In a similar fashion, here, measures of coherence such as the relative entropy of coherence~\cite{relentropy} serve the analogous purpose.

We now proceed to augment the discussion on causality to take into account the uncertainty of the moment of measurement. If the events have non-negligible temporal uncertainty and $\cal S$ is non-stationary, but the experimentalist lacks access to information recorded by the timers, we should trace the timers out from ${\hat \rho}_\varepsilon$. Here one can confirm that coherence still only exists for $\varepsilon={\rm {TL}}$. See Eqs.~(\ref{apTL}) and~(\ref{apSL}) in the Appendix B for a visualization.

Finally, for the general event in Eq.~(\ref{rhoe}), where the experimentalist learns what the timers register, we consider the following correlation function to differentiate SLE from TLE,
\begin{eqnarray}\label{corre}
C^\varepsilon_{AB}({\hat T})={\big \langle} ({\hat T}_A-\langle {\hat T}_A \rangle_\varepsilon) ({\hat T}_B-\langle {\hat T}_B \rangle_\varepsilon){\big \rangle}_\varepsilon.
\end{eqnarray}
$C^\varepsilon_{AB}({\hat T})$ is computed for the state of Eq.~(\ref{rhoe}) in Appendix C. We verify $C^\varepsilon_{AB}({\hat T}) > 0$ for TLE and is zero for SLE discussed above. For example, in  two causally connected measurements, if the first measurement occurs earlier (later) than its expected time of occurrence, the second one will also (on average) happen earlier (later), resulting in a positive $C^{\rm TL}_{AB}({\hat T})$.

\textit{QI theory of events and determinism in QM}--- In spatially entangled systems, one can quantify the information gained about one subsystem given some measurement of the other. Our focus on detectors and timers allows us to similarly quantify, for causally connected events, how much information we know about the future (spatial analogue of the unobserved system) given some information about the past (analogous to observed spatial subsystem). For SLE, the standard interpretation~\cite{book1, book2, Vedral2,Holevo} holds.

Again, for simplicity, let us consider the case where $\chi$ is very short allowing us to neglect the timers. To analyze equations Eqs.~(\ref{rhoSL}) and (\ref{rhoTL}) in the context of QI ~\cite{Vedral2,Holevo}, we rewrite them in the format
\begin{eqnarray}
\label{rhoee}
{\hat \rho}_{\varepsilon}=\sum_{\alpha_B}~p^{\varepsilon}_{\alpha_B}~{\hat \sigma}^\varepsilon_{A,\alpha_B}\otimes
|\alpha_B\rangle_{B}{\langle}\alpha_B|,\end{eqnarray}
where, for SLE, we have
\begin{eqnarray}
\label{rhoSL2}
{\hat \sigma}^{{\rm S}{\rm L}}_{A,\alpha_B}=\frac{1}{p^{{\rm S}{\rm L}}_{\alpha_B}}~\sum_{\alpha_A}~{\big|}\langle \alpha_A,\alpha_B|\varphi(t_0)\rangle_{\cal S}{\big |}^2 ~|\alpha_A\rangle_{A}{\langle}\alpha_A|,~~~~
\end{eqnarray}
with $p^{{\rm S}{\rm L}}_{\alpha_B}=\sum_{\alpha_A}~{\big|}\langle \alpha_A,\alpha_B|\varphi(t_0)\rangle_{\cal S}{\big |}^2$, and for TLE, we have
\begin{eqnarray}
\label{rhoTL2}
{\hat \sigma}^{\rm {TL}}_{A,\alpha_B}=|\lambda_{\alpha_B}\rangle_A{\langle}\lambda_{\alpha_B}|,\end{eqnarray}
with $|\lambda_{\alpha_B}\rangle_A=1/{\sqrt{p^{{\rm T}{\rm L}}_{\alpha_B}}}\sum_{\alpha_A}\langle \alpha_B|{\hat U}|\alpha_A\rangle\langle \alpha_A|\varphi(t_0)\rangle_{\cal S}|\alpha_A\rangle_A$
and $p^{{\rm T}{\rm  L}}_{\alpha_B}=\sum_{\alpha_A}~|\langle \alpha_B|{\hat U}|\alpha_A\rangle|^2~|\langle \alpha_A|\varphi(t_0)\rangle_{\cal S}|^2$. Notice that $p^\varepsilon_{\alpha_B}$ is the probability of ${\cal D}_B$ measuring $\alpha_B$ regardless of the value of $\alpha_A$ measured by ${\cal D}_A$.

Now, recall Eq.~(\ref{rhoee}) and  consider that Alice observes ${\cal D}_A$ in her laboratory to learn the state of $\cal S$ while being agnostic to the causal nature of the events. As long as ${\cal D}_A$ does not undergo decoherence, Alice can measure ${\cal D}_A$ in a basis different from $\{|\alpha_A\rangle_A\}$, which stores information about the states $\{|\alpha_A\rangle\}$ of $\cal S$. Assuming that Bob always measures ${\cal D}_B$ in the basis $\{|\alpha_B\rangle_B\}$, we observe from Eq.~(\ref{rhoee}) that Alice can infer the eigenvalue $\alpha_B$ measured by Bob by trying to discriminate between the different states ${\hat \sigma}^{\varepsilon}_{A,\alpha_B}$ of her detector. In particular, for two level systems $\alpha_B=1,2$, Alice's highest probability of success is given by
$
\label{Disc}
p_{\rm suc}= 1/2 ( 1 + \mathrm{Tr}|p_2^\varepsilon\hat{\sigma}_{A,1}^\varepsilon - p_1^\varepsilon\hat{\sigma}_{A,2}^\varepsilon|)
$
according to Helstrom's discrimination criteria. The projection operators with the highest probability of success satisfy constraints outlined in \cite{Holevo}. Note that we are employing QI tools to characterize the detectors' states, instead of the usual application on the systems under measurement. As we will discuss below, this approach allows us to extend the existing information measures and QI protocols to TLE.

Reference~\cite{Vedral2} provides a way to compute the largest amount of information gained by Alice about Bob's observation. This quantity is a classical correlation function which when employed to our case yields $C_A({\hat \rho}_B^\varepsilon)={\rm max}_{{{\hat M}^{\dagger}_{A,i} {\hat M}_{A,i}}} \{S({\hat \rho}_B^\varepsilon)-\sum_{i}~p_{i}~S({\hat \rho}_{B,i}^\varepsilon)\}$,
where ${\hat \rho}_B^\varepsilon={\rm Tr}_A{\hat \rho}_{\varepsilon}$, ${\hat \rho}_{B,i}^\varepsilon={\rm Tr}_A({\hat M}_{A,i}{\hat \rho}_{\varepsilon}{\hat M}^{\dagger}_{A,i})/p_i$ is the state of ${\cal D}_A$ after the outcome $i$ is observed by Alice, and $p_i={\rm Tr}_{AB}({\hat M}_{A,i}{\hat  \rho}_\varepsilon{\hat M}^{\dagger}_{A,i})$. In~\cite{Vedral2}, $C_A$ was proposed only for SLE, in which its evaluation involved the state of the system under detection [i.e., $C_A({\hat \rho}_{{\cal S}_B})$] and not the states of the detectors as considered here. In SLE, the primary difference between $C_A$ here and that from Ref.~\cite{Vedral2} is that we compute the correlations between measurement records of subsystems of $\cal S$ [$C_A({\hat \rho}_B^\varepsilon)$] and not correlations between subsystems of $\cal S$. We give an example of Eq.~(\ref{rhoee}) and $C_A$ for SLE and TLE in Appendix E.

For SLE, consider  $|\varphi(t_0)\rangle_{\cal S}=\sum_\alpha c_\alpha|\alpha_A\rangle|\alpha_B\rangle $. Thus, Eq.~(\ref{rhoee}) becomes ${\hat \rho}_{{\rm S}{\rm L}}=\sum_{\alpha} |c_\alpha|^2 |\alpha_A\rangle_{A}{\langle}\alpha_A|\otimes
|\alpha_B\rangle_{B}{\langle}\alpha_B|$. Thus clearly the best measurement is in $\{|\alpha_A\rangle\}_A$ of ${\cal D}_A$, which results in $C_A({\hat \rho}_B^{\rm {SL}})=-\sum_\alpha |c_\alpha|^2 {\rm log}|c_\alpha|^2$. Analogously for TLE this happens when, e.g., the initial condition is $|\varphi(t_0)\rangle_{\cal S}=\sum_\alpha c_\alpha|\alpha\rangle$, ${\hat U}=\unit$, and the same observable is measured twice. 

We present an interpretation of Alice's prediction for TLE in the following. Recall that for TLE, $\mathcal{D}_A$ and $\mathcal{D}_B$ in Eq.~(\ref{rhoee}) record information about the same system $\cal S$ at an earlier and later time, respectively. Thus, we can say that Alice's guess about the state measured by Bob is a prediction of a future measurement of $\cal S$, which refers to the knowledge Bob gains while measuring ${\cal D}_B$ in the basis $\{|\alpha_B\rangle_B\}$. Recall that ${\cal D}_B$ records the value $\alpha_B$ of $\cal S$. In this manner, by fixing the evolution ${\hat U}$ and the observables of $\cal S$ measured by ${\cal D}_A$ and ${\cal D}_B$, Alice can predict the future event of $\cal S$  (measurement performed by ${\cal D}_B$ and checked later by Bob) by measuring ${\cal D}_A$ (by observing a past record) in the basis $\{|\lambda_{\alpha_B}\rangle_A\}$, with a probability of success given by $p_{\rm suc}$. Unlike ${\hat \sigma}^{{\rm S}{\rm L}}_{A,\alpha_B}$, ${\hat \sigma}^{{\rm T}{\rm L}}_{A,\alpha_B}$ is a pure state and hence $p_{\rm suc}$ for a two-level system becomes the original Helstrom's discrimination $
p_{\rm suc}=1/2(1+\sqrt{1-4 p_{1} p_{2}|{_A\langle}\lambda_{1}|\lambda_{2}\rangle_A|^2})$ \cite{Helstrom}. In general the states of the set $\{|\lambda_{\alpha_B}\rangle_A\}$ are not orthogonal.

If ${_A\langle}\lambda_{\alpha_B}|\lambda_{\alpha'_B}\rangle_A = \delta_{\alpha_B, \alpha'_B}$, $p_{\rm suc}=1$ and Alice can predict deterministically what Bob will learn, or learned, about $\cal S$. Taking a different perspective from the above point of view, let us consider that the observable measured by $\cal{D}_A$ is not predetermined. So, given ${\hat U}_{\cal S}$ and the basis measured by ${\cal D}_B$, Alice can always set ${\cal D}_A$ to measure an observable of $\cal S$ prior to Bob's measurement such that ${_A\langle}\lambda_{\alpha_B}|\lambda_{\alpha'_B}\rangle_A = \delta_{\alpha_B, \alpha'_B}$, thus having a deterministic prediction. This striking feature is similar to classical determinism if we assume that the information stored by ${\cal D}_A$ defines the initial condition of $\cal S$. This behaviour comes from the fact that ${\cal D}_A$ and ${\cal D}_B$ become maximally correlated for TLE when ${_A\langle}\lambda_{\alpha_B}|\lambda_{\alpha'_B}\rangle_A = \delta_{\alpha_B, \alpha'_B}$ [see Eqs.~(\ref{rhoee}) and (\ref{rhoTL2})].

A future investigation would be the application of QI tools to the timers, i.e., to the case where Alice and Bob have access to when the events take place. Lastly, it is worth remarking that the temporal superpositions of ${\hat \rho}_{\varepsilon}$ can naturally give rise to indefinite causal orders~\cite{progress}, similar to those seen via quantum gates~\cite{Chir,Chir2,Felce}. 

\section{Acknowledgements}
EOD acknowledges financial support from Conselho Nacional de Desenvolvimento Científico e Tecnológico (CNPq) through its program  09/2020 (Grant No. 315759/2020-8) and Coordenação de Aperfeiçoamento de Pessoal de Nível Superior (CAPES) through its program GPCT - 17/2016 (Grant No. 88887.312745/2018-00). VV thanks the National Research Foundation, Prime Minister’s Office, Singapore, under its Competitive Research Programme (CRP Award No. NRFCRP14-T262014-02) and administered by Centre for Quantum Technologies, National University of Singapore. This publication was made possible through the support of the ID 61466 grant from the John Templeton Foundation, as part of the The Quantum Information Structure of Spacetime (QISS) Project (qiss.fr). The opinions expressed in this publication are those of the authors and do not necessarily reflect the views of the John Templeton Foundation.

\begin{appendix}
\section{Appendix A. A simple example for the computation of the measurement probabilities}
\label{ap0}
For the simplest case where $\delta p_{\s(k)}^j=\delta p_j\ll 1$, the relation $|\chi_j(t_{\s(k)})|^2~dt=|\tilde{\chi}_j(t_{(k+1)},t_{\s(k)})|^2$ becomes
\begin{eqnarray}
\nonumber
\delta p_{j}(1-k\delta p_{j}) 
&\;\;\;\;\approx |\chi_{j}(t_{(k)})|^2~\delta t
\\\Rightarrow (\delta p_j/\delta t){\rm e}^{-k\delta p_j}
&\approx |\chi_{j}(t_{(k)})|^2
\end{eqnarray}
By defining $\delta p_j=\gamma_j \delta t$ and recalling that $t_{(k)}-t_0=k\delta t$, we have,
\begin{eqnarray}\label{exp}
\chi_j(t_j)=
\begin{cases}&\sqrt{\gamma_j}{\rm e}^{-\gamma_j (t_j-t_0)/2} \;\;\;\;
\mathrm{for} \;\;t_j\geq t_0\\
&\;\;\;\;\;\;\;\;\;\;0 \;\;\;\;\;\;\;\;\;\;\;\;\;\;\;\;\;\; \mathrm{otherwise}.
\end{cases}
\end{eqnarray}

Equation~(\ref{exp}) is the time probability amplitude for the measurements of a decay photon. To understand this exponential dependence of $\chi$, first recall that from standard calculations in QM, the probability for an atom initially excited measured at time $t_j$ to be found in the ground state is $\propto 1- {\rm e}^{-\gamma_j (t_j-t_0)/2}$. Then, as it is an irreversible process, the time probability density of decay is simply $\propto -d({\rm e}^{-\gamma_j (t_j-t_0)/2})/dt$, i.e., $\propto |\chi_j(t_j)|^2$ above. In this specific example, the computation of $|\chi_j(t_j)|^2$ is straightforward since the decay is not affected by both the photon measurement and register of the time at which this detection happens. 

\section{Appendix B. Fuzzy time events without observing the timers}
\label{ap}
We considered events of a long duration where it was necessary to include a timer to record the instance of measurement. Here we consider the situation where the experimentalist does not have access to the information stored in the timers for which we obtain
\begin{eqnarray}\label{apTL}
    \hat{\rho}_{\rm {TL}} =\sum_{\alpha_A,\alpha_B} \sum_{\alpha'_A}~\int_{t_0}^\infty\int_{t_A}^\infty~dt_A dt_B~ |\chi_B(t_B|t_A)|^2 |\chi_A(t_A)|^2 &&\nonumber\\
    \langle \alpha_B| \hat{U}_{\cal{S}}(t_B,t_A) |\alpha_A\rangle ~\langle \alpha_A|\hat{U}_{\cal S}(t_A,t_0) |\psi(t_0)\rangle_{\cal S} &&\nonumber\\
    {_{\cal S}\langle} \psi(t_0)| \hat{U}_{\cal S}^\dagger(t_A,t_0)|\alpha'_A\rangle \langle \alpha'_A| \hat{U}_{\cal S}^\dagger (t_B,t_A) |\alpha_B\rangle &&\nonumber\\
    |\alpha_A\alpha_B\rangle_{AB}\langle\alpha'_A\alpha_B|.~~~~~~~~
\end{eqnarray}
By rewriting this equation in the notation of Eq.~(\ref{rhoee}), with $p^\varepsilon_{\alpha_B}~{\hat \sigma}^\varepsilon_{A,\alpha_B}= \int dt_B \int dt_A~{ p}'^{\varepsilon}_{\alpha_B}(t_A,t_B)~{\hat \rho}'^{\varepsilon}_{\alpha_B}(t_A,t_B)$, we obtain
\begin{eqnarray}
    {p}'^{\rm {TL}}_{\alpha_B}(t_A,t_B) = \sum_{\alpha_A} |\langle \alpha_B| \hat{U}_{\cal S}(t_B,t_A) |\alpha_A\rangle|^2\nonumber\\
    |\langle \alpha_A|\hat{U}_{\cal{S}}(t_A,t_0) |\psi(t_0)\rangle_S|^2
\end{eqnarray}
and
\begin{eqnarray}
    {\hat \rho}'^{\rm{TL}}_{\alpha_B}(t_A,t_B) = |\lambda_{\alpha_B}(t_A,t_B)\rangle \langle \lambda_{\alpha_B}(t_A,t_B)|,
\end{eqnarray}
with
\begin{eqnarray}
|\lambda_{\alpha_B}(t_A,t_B)\rangle = 
\frac{1}{\sqrt{{p}'^{\rm {TL}}_{\alpha_B}(t_A,t_B)}}\sum_{\alpha_A} \langle \alpha_B| \hat{U}_{\cal S}(t_B,t_A) |\alpha_A\rangle \nonumber\\
\langle \alpha_A|\hat{U}_{\cal S}(t_A,t_0)|\psi(t_0)\rangle_S|\alpha_A\rangle.~~~~~
\end{eqnarray}
Similarly, for SLE, we have
\begin{eqnarray}\label{apSL}
\hat{\rho}_{\rm {SL}} =\sum_{\alpha_A,\alpha_B}~\int_{t_0}^\infty\int_{t_0}^\infty~dt_A dt_B~  |\chi_B(t_B)|^2 |\chi_A(t_A)|^2 \nonumber\\
\big|\langle \alpha_A \alpha_B|\hat{U}_{{\cal S}_A}(t_A,t_0) \otimes 
 \hat{U}_{{\cal S}_B}(t_B,t_0)|\psi(t_0)\rangle_{\cal S}\big|^2\nonumber\\
 |\alpha_A \alpha_B\rangle_{AB} \langle \alpha_A \alpha_B|~~~
\end{eqnarray}
where we assume the separable evolution of the system $\hat{U}_{{\cal S}}(t,t_0)=\hat{U}_{{\cal S}_A}(t,t_0) \otimes 
 \hat{U}_{{\cal S}_B}(t,t_0)$  that stops after their measurements, i.e., $\hat{U}_{{\cal S}_j}(t,t_j)=0$ with $t_j$ being the detection time of ${\cal S}_j$ .

Now the notation of Eq.~\eqref{rhoee} yields
\begin{eqnarray}
    &&{p}'^{\rm {SL}}_{\alpha_B}(t_A,t_B)= \nonumber\\ 
    &&\sum_{\alpha_A} \big|\langle \alpha_A \alpha_B|\hat{U}_{{\cal S}_A}(t_A,t_0) \otimes 
 \hat{U}_{{\cal S}_B}(t_B,t_0)|\psi(t_0)\rangle_{\cal S}\big|^2~~~~~~~
\end{eqnarray}
and
\begin{eqnarray}
&&{\hat \rho}'^{\rm {SL}}_{\alpha_B}(t_A,t_B) =\frac{1}{{p}'^{\rm {SL}}_{\alpha_B}(t_A,t_B)} \nonumber\\
&&\sum_{\alpha_A} \big|\langle \alpha_A \alpha_B|\hat{U}_{{\cal S}_A}(t_A,t_0) \otimes 
 \hat{U}_{{\cal S}_B}(t_B,t_0)|\psi(t_0)\rangle_{\cal S}\big|^2|\alpha_A\rangle \langle\alpha_A|.\nonumber\\
\end{eqnarray}
Note that in the SL case we need to consider three terms corresponding to the order of the measurements. Furthermore, from the above results it is evident that the coherence is retained in the state describing TLE while it is absent in SLE.

\section{Appendix C. The behaviour of the causality measure}
\label{ap2}

We will study the properties of the causality correlation in Eq. \eqref{corre} function by rewriting it as
\begin{equation}
   C^\varepsilon_{AB}({\hat T})= \langle \hat{T}_A \hat{T}_B \rangle_\varepsilon - \langle \hat{T}_A \rangle_\varepsilon\langle \hat{T}_A \rangle_\varepsilon.
\end{equation}
For TLE,
\begin{eqnarray}
    \nonumber
   C^{\rm {TL}}_{AB}({\hat T})&=&\int_{t_0}^\infty\int_{t_a}^\infty~dt_A dt_B~ t_A t_B |\chi_B(t_B|t_A)|^2 |\chi_A(t_A)|^2 
    \\\nonumber
    &-&\int_{t_0}^\infty\int_{t_a}^\infty~dt_A dt_B~ t_A  |\chi_B(t_B|t_A)|^2 |\chi_A(t_A)|^2 
    \\
    &\times& \int_{t_0}^\infty\int_{t_a}^\infty~dt_A dt_B~ t_B |\chi_B(t_B|t_A)|^2 |\chi_A(t_A)|^2.\nonumber\\
\end{eqnarray}
Finally, by writing the above expression in the following succinct notation, we conclude that
\begin{eqnarray}
\label{CCFINEQ}
    \nonumber
   &&C^{\rm {TL}}_{AB}({\hat T})= \int_{t_0}^\infty~dt_A t_A |\chi_A(t_A)|^2~ \mathrm{E}(\hat{T}_B|t_A)
    \\\nonumber
    &&-\int_{t_0}^\infty~dt_A ~ t_A  |\chi_A(t_A)|^2
    \times \int_{t_0}^\infty~dt_A  |\chi_A(t_A)|^2~ \mathrm{E}(\hat{T}_B|t_A)\nonumber\\
    &&\equiv
    \langle fg \rangle_{\chi_A} - \langle f \rangle_{\chi_A} \langle g \rangle_{\chi_A}
    \geq 0
\end{eqnarray}
where $\mathrm{E}(\hat{T}_B|t_A) = \int_{t_A}^\infty~dt_B t_B  |\chi_A(t_B|t_A)|^2 > t_A~\forall t_A$, $f=t_A$, and $g=\mathrm{E}(\hat{T}_B|t_A)$. In the final line of Eq.~\eqref{CCFINEQ}, we apply the Chebyshev-Harris inequality~\cite{Harris} for the distribution $\chi_A$. 

In SLE with a separable $\chi_{\rm{SL}}(t_A,t_B)=\chi_A(t_A)\chi_B(t_B)$, the measure is trivially 0 as the probability amplitude of the second event is not conditional upon the moment of occurrence of the first event. In this case the normalization of the probability amplitude coupled with the above independence yields straightforwardly $C^{\rm {SL}}_{AB}({\hat T})=0$.

\section{Appendix D. An event based approach to Bell's inequalities}\label{ap2}
Bell’s theorem demonstrates that under a wide class of conditions pertaining to locality and causality, no other theory can completely reproduce the probabilistic predictions of quantum mechanics. A particular manifestation of Bell’s theorem is the fact that entangled states i.e., quantum states of composite systems that cannot be expressed as a (mixtures of) product state, violate Bell's inequalities (BIs). 

Since quantum correlations can be accessed by Bell like inequalities, we investigate the extension of BIs to events in our formalism. Traditionally, for a bipartite system ${\cal S}={\cal S}_A+{\cal S}_B$ living in $\mathcal{H}_{{\cal S}_A} \otimes \mathcal{H}_{{\cal S}_B}$, the correlation functions of BIs are given by ${\rm Tr}({\hat \rho}_{\cal S}{\hat A}\otimes{\hat B})$, i.e., they involve the expectation values of observables ${\hat A}$ and ${\hat B}$ of a spatially separated bipartite system ($\cal S$) chosen to be measured by Alice (A) and Bob (B). Although BIs have been applied to TLE in Ref.~\cite{Vedral3}, due to the space-time asymmetry of QM, the calculation of correlations via ${\rm Tr}({\hat \rho}_{\cal S}{\hat A}\otimes{\hat B})$ limits BIs to spacelike measurements (or SLE).

In this work, on the other hand, as we focus on the detectors' states and not on the states of $\cal S$, BIs are calculated identically for SLE and TLE. For a pair of observables ${\hat A}$ and ${\hat B}$, the correlation function is given by ${\rm Tr}({\hat \rho}_\varepsilon {\hat A}_{{\cal D}_A}\otimes{\hat B}_{{\cal D}_B})$, with ${\hat A}_{{\cal D}_A}$ (${\hat B}_{{\cal D}_B}$) taken to be the observable of ${\cal D}_A$ (${\cal D}_B$) whose eigenstates record information of $\hat A$ ($\hat B$). Here, $\hat A$ ($\hat B$) is either the observable of ${\cal S}_A$ (${\cal S}_B$) or the observable of the first (second) measurement of $\cal S$. For example, in CHSH inequality, one has to compute four different density matrices ${\hat \rho}_\varepsilon$, one for each pair of observables that Alice and Bob have chosen for ${\cal D}_A$ and ${\cal D}_B$ to measure. This procedure is faithful to the standard CHSH inequality on a bipartite state and is extendable to subsequent measurements on the same system when we utilize the event state description provided in \eqref{rhoe}.

\section{Appendix E. Predicting a future event looking at a past record}\label{ap2}
A simple example of the deterministic feature of events is visualized by considering a spin-half system $\cal S$ initially in the state $|\varphi(t_0)\rangle_{\cal S}=|+_x\rangle$, evolving according to ${\hat U}=\exp\{i\pi \sigma_x/4\}$, and subsequently measured in the ${\hat S}_y$ basis by ${\cal D}_B$. By knowing this configuration, Alice can predict the outcome measured by ${\cal D}_B$ (seen by Bob) by setting ${\cal D}_A$ to previously measure $\cal S$ in the basis ${\hat S}_z$ so that $|\lambda_{+_y}\rangle_A=|+_z\rangle_A$ and $|\lambda_{-_y}\rangle_A=|-_z\rangle_A$. Thus, Eq.~(\ref{rhoee}) becomes 
$
{\hat \rho}_{{\rm T}{\rm L}} = 1/2( |+_z\rangle_{A}{\langle}+_z|\otimes
|+_y\rangle_{B}{\langle}+_y|
+|-_z\rangle_{A}{\langle}-_z|\otimes
|-_y\rangle_{B}{\langle}-_y|),
$
where we verify that, before (after) Bob's observation, if Alice measures $|\pm_z\rangle_A$, she knows Bob will measure (measured) $|\pm_y\rangle_B$. In this manner, Alice looks at a past record in such a way as to precisely learn about a future record. Finally, in Appendix B we show that if  the  events have non negligible $\chi$ but Alice and Bob lack access to the timers, the previous discussion about Alice's predictions still hold.

\end{appendix}


\begin{thebibliography}{99}
\bibitem{footnote1}
For intrinsic uncertainty in the instant of measurement, it suffices if the system travels a certain distance until it reaches the apparatus (arrival time problem) or if the interaction time between the detector and system is not short enough to neglect the evolution of the system in the absence of measurements. Note there is still no consensus on the best strategy to determine arrival times, and that our formalism is applicable to compute it.
\bibitem{footnote2}In this context, the branches of $|\psi(t_{\s (1)})\rangle$ in which the particle is not measured can arise from both the uncertainty of the particle's arrival time at the detector and the evolution of the detector until the click. The click can simply be the transition from the detector's initial state to an orthogonal one, when the timer stops running.
\end{thebibliography}
\end{document}